# Low-energy spin precession in the molecular field of a magnetic thin film

Christopher Vautrin[1], Daniel Lacour[1], Coriolan Tiusan[2*], Yuan Lu[1], François Montaigne[1], Mairbek Chshiev[3], Wolfgang Weber[4], Michel Hehn[1*]

[1] Institut Jean Lamour, Campus Artem, 2 allée André Guinier, BP50840 - F-54011 Nancy
[2] Center of Superconductivity, Spintronics and Surface Science (C4S), Technical University of Cluj-Napoca, 400114 Romania
[3] Univ. Grenoble Alpes, CEA, CNRS, SPINTEC, 38000, Grenoble, France
[4] Institut de Physique et Chimie des Matériaux de Strasbourg, UMR 7504 CNRS, Université de Strasbourg, 23 Rue du Loess, BP 43, 67034 Strasbourg Cedex 2, France

*Corresponding author: michel.hehn@univ-lorraine.fr, coriolan.tiusan@phys.utcluj.ro

Electronic spin precession and filtering are measured in the molecular field of magnetic thin films. The conducted lab-on-chip experiments allow injection of electrons with energies between 0.8 and 1.1 eV, an energy range never explored up to now in spin precession experiments. While filtering angles agree with previous reported values measured at much higher electron energies, spin precession angles of 2.5° in CoFe and 0.7° in Co per nanometer film thickness could be measured which are 30 times smaller than those previously measured at 7 eV. Band structure effects and layer roughness are responsible for these small precession angle values.

Since the discovery of Giant Magneto-Resistance (GMR) in 1988[1,2] the Spintronics research field has become the ground of intense investigations. Accompanying the growth of fundamental knowledge on spin transport in solid-state devices, numerous proposals for applications have emerged (see for instance[3] and [4]). Among those, some have already hit the market as hard drive read heads, magnetic fields sensors and spin transfer torque based magnetic random access memories (STT-MRAMs). Despite the apparent maturity of the field, fundamental points remain to be clarified[5]. In particular, very little is known about the electronic spin behavior in its out-of-equilibrium state (*i.e.* beyond the Fermi sea) even if it is acknowledged to be a source of spin-transfer torque[6].

When a beam of electrons with an initial spin polarization vector $\boldsymbol{P_0}$ is injected into a region of space where a magnetic field $\boldsymbol{H}$ is present, the polarization vector $\boldsymbol{P}$ will exhibit a precessional motion around the magnetic field. Two angles can then be defined: the filtering angle, $\theta$, that describes the reorientation of $\boldsymbol{P}$ towards $\boldsymbol{H}$ and the precession angle, $\varepsilon$, that describes the precession of $\boldsymbol{P}$ around $\boldsymbol{H}$ (figure 1 in which the blue arrow is the field $\boldsymbol{H}$). The precession frequency is given by the Larmor frequency $\omega_L = \gamma H$ with $\gamma \simeq 1{,}7 \cdot 10^{11}\ \text{rad}\ \text{s}^{-1}\ \text{T}^{-1}$ the gyromagnetic ratio. If the incident polarized electrons are considered to move at a typical speed of $2 \cdot 10^6\ \text{ms}^{-1}$ (close to the Fermi velocity of many metallic elements), a precession angle per µm and per Tesla, $\tilde{\varepsilon}$, of about 0,17 $\text{rad}\ \text{T}^{-1}\text{µm}^{-1}$ is expected. Large precession angles can thus be achieved either by a short travel distance in a strong magnetic field or by a long travel distance in a small magnetic field. This latter scheme was used in metals by Jedema et al[7] as well as in semiconductors by I. Appelbaum et al[8] and by D. D. Awschalom et al[9], while the first strategy was employed by Oberli et al[10] in their free-electron beam experiments. By injecting a spin polarized electron beam into a magnetic layer, the so-called molecular field of a ferromagnetic layer is estimated to be of the order of several 100 T to 1000 T. Oberli et al achieved experimentally precession angles of several tens of degrees per nanometer. Unfortunately, measurements with a free electron beam at electron energies (with respect to the Fermi level) below the vacuum level (4-5 eV for ferromagnetic metals, such as Co and Fe) are not possible. However, this is exactly the energy range of interest for all spintronics applications (typical bias voltages applied to tunnel junctions are 1 to 2 V). Therefore, we have conducted lab-on-chip experiments allowing us to measure at these low energies the spin precession induced by the molecular field of thin ferromagnetic layers.

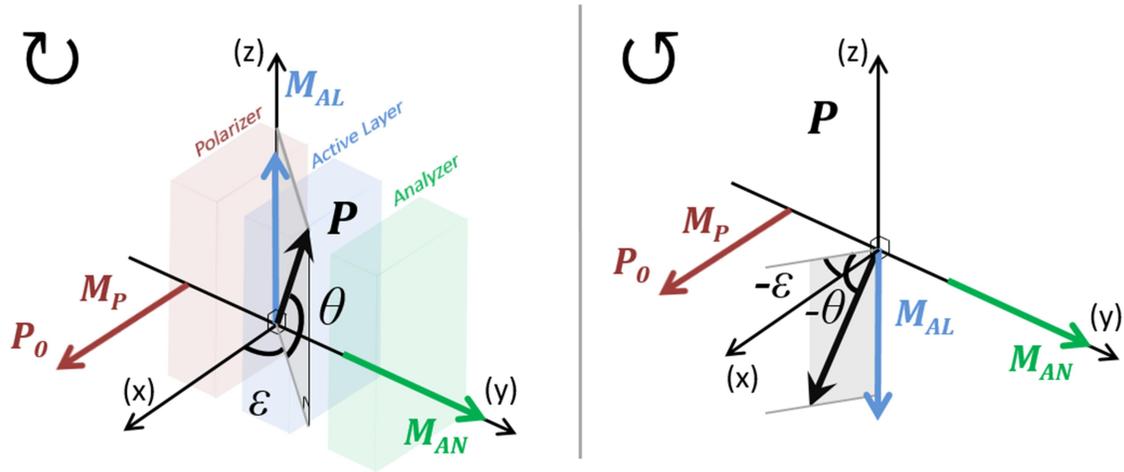

*__Figure 1:__ Angles of precession and filtering of the spin polarization vector of an electron beam injected into a magnetic layer with magnetization $M_{AL}$ or a field H oriented along $M_{AL}$. See text for further details.*

Three elements are essential to perform such experiments: a spin polarizing layer, an active precession layer and an analyzing layer. In order to study the dependence of the precession angle as a function of electron energy, the lab-on-chip has to host an electronic device that allows varying the injection energy of the electrons. In this work the electron injection is accomplished by a magnetic tunnel junction (figure 2). After having aligned the spin polarization of the injected electrons within the polarizing layer along its magnetization direction, the electrons are transported via the tunnel effect through the MgO barrier. As the tunnel transport is spin conservative, the spin polarization of the electrons arriving in the active layer is perpendicular to the active's layer magnetization direction and will consequently precess around it during the electron's propagation. Changing the bias voltage across the tunnel barrier varies the injection energy in the active layer and thus allows a spectroscopic analysis of the precession angle. This angle is analyzed through the GMR effect occurring in the active layer / Cu / analyzer spin valve (blue/orange/green rectangles in Figure 2). Note that the analyzer magnetization is orthogonal to the magnetization of both the active and the polarizing layers. Finally, one last key ingredient to the precession angle analysis is a Schottky diode (figure 2). It allows a dual analysis. First of all, it ensures that the collected electrons in the semiconductor have always an energy higher than 0.7 eV (the height of the Schottky barrier) after having passed the spin valve. Thus, only the spin precession of hot electrons is analyzed, while all thermalized electrons are reinjected by the tunnel barrier through the spinvalve's electron recovery circuit. Second, it defines an acceptance cone with an opening angle θc of only 4.5° at E = 1 eV at the Schottky interface due to the conservation of the momentum parallel to the Cu/Si interface (see additional material). This angular wave vector filtering effect is reinforced by another angular filtering taking place during the tunnelling process. Thus, for an injection energy of around 0.7 eV, the collected electrons in silicon have been transported across the spin valve in an almost ballistic manner and practically perpendicular with respect to the multilayer interfaces.

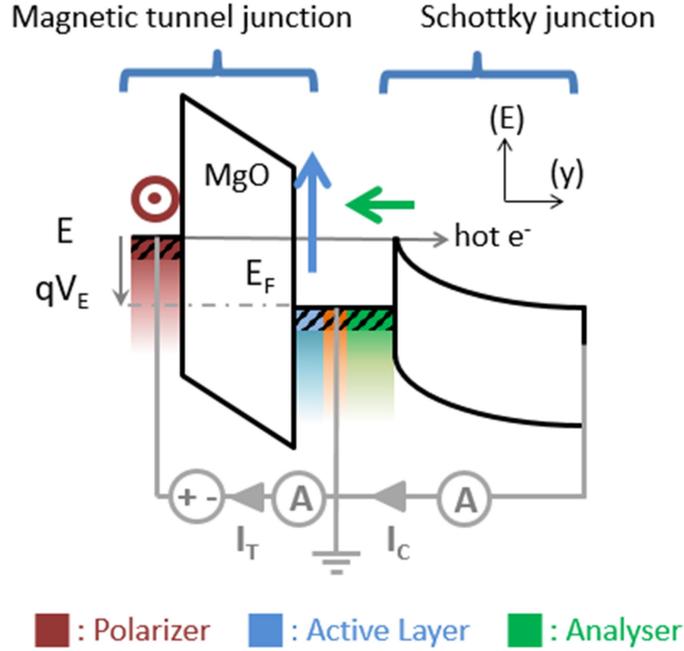

***Figure 2:*** *All solid state device based on a magnetic tunnel transistor that allows measuring the spin precession at low energies in the molecular field of a thin magnetic layer. See text for further details.*

A typical stack used in this study is as follows: Pt(5)/IrMn (7.5)/Co(2)/Ta(0.5)/CoFeB(2)/MgO(2.5)/X(y)/Cu(3.5)/[Ni(0.6)/Co(0.2)]x5/Ni(0.6)/Cu(5)/Ta(1)/ Cu(5)//Si(100), where numbers in brackets indicate the layer thicknesses in nm. The multilayer is grown by sputtering on a HF cleaned Si substrate (see method section for supplementary details). The CoFeB layer is the polarizer and X is the active layer. The [Ni(0.6)/Co(0.2)]x5/Ni(0.6) multilayer represents the analyzer. To determine a precession angle in our lab-on-chip experiment it is required to stabilize the aforementioned three-dimensional magnetic configuration. The way to obtain a crossed configuration of spin valve magnetizations has been reported in a previous study[11]. The crossed configuration in the Pt/IrMn/Co/Ta/CoFeB/MgO/X(y) tunnel junction is obtained by establishing an exchange bias field at the IrMn/Co interface to set the magnetization easy axis of the polarizer along the x-direction. The multilayer deposition is followed by an annealing process under an applied field to initiate the exchange bias field and by four steps of optical lithography to define the electrical contacts on the different layers of interest (see method section for supplementary details).

The collected hot electron current $I_C$ can be expressed as $I_C = I_C^\perp (1 + MC^\perp \boldsymbol{P} \cdot \boldsymbol{M_{An}})$ where $\boldsymbol{P}$ is the hot electron spin polarization vector after propagation through the active layer and $\boldsymbol{M_{An}}$ indicates a unit vector pointing along the magnetization direction of the analyzer[12]. $MC^\perp$ is the magneto-current ratio defined as $\frac{I_C^\parallel - I_C^\perp}{I_C^\perp}$ where is $I_C^\parallel$ and $I_C^\perp$ are the collected currents when $\boldsymbol{P}$ and $\boldsymbol{M_{An}}$ are parallel or perpendicular to each other, respectively. Since we plan a spectroscopic analysis, the tunnel junction bias voltage ($V_E$) will be changed and so will the injected current. It is then convenient to normalize the collected current by the injected one. This defines the transfer ratio as:

$$TR = TR^{\perp}(1 + MC^{\perp} \boldsymbol{P} \cdot \boldsymbol{M_{An}}),$$

which could also be expressed as $TR^{\perp}[1 + MC^{\perp} P_0 \cos(\theta) \sin(\varepsilon)]$. The $\cos(\theta)\sin(\varepsilon)$ product containing our angles of interest can thus be nicely obtained experimentally by measuring the transfer ratio in three different magnetic configurations: parallel ∥ [13], clock wise ↻ and counter clock wise ↺ as illustrated in figure 1. The three transfer ratios $TR^{\parallel}$, $TR^{\circlearrowright}$ and $TR^{\circlearrowleft}$, can be expressed as $TR^{\parallel} = TR^{\perp}(1 + MC^{\perp} P_0)$ and $TR^{\circlearrowright/\circlearrowleft} = TR^{\perp}[1 \pm MC^{\perp} P_0 \cos(\theta) \sin(\varepsilon)]$. As a result, the $\sin(\varepsilon)\cos(\theta)$ product as a function of experimentally available quantities writes as:

$$\sin(\varepsilon)\cos(\theta) = \frac{TR^{\circlearrowright} - TR^{\circlearrowleft}}{2TR^{\parallel} - (TR^{\circlearrowright} + TR^{\circlearrowleft})}.$$

Measurements of the three aforementioned $TR$ vs. $V_E$ performed on the sample having an active layer composed of a 1 nm thick CoFeB film are reported in figure 3a. They provide clear evidence of a precessional effect in the CoFeB layer since $TR^{\circlearrowright}$ and $TR^{\circlearrowleft}$ are not superimposed. Precession angles $\varepsilon$ obtained for different filtering angles $\theta$ ranging from 0 ° to 85 ° are shown in figure 3b. One must note that the obtained values are always smaller than the ones reported by Weber et al[14] and this even if strong filtering effects are considered. Increasing $\theta$ towards 90 ° naturally increases $\varepsilon$ towards Weber's values but such a high spin filtering effect is not expected for such a thin magnetic layer. In order to determine $\varepsilon$, the determination of $\theta$ is mandatory.

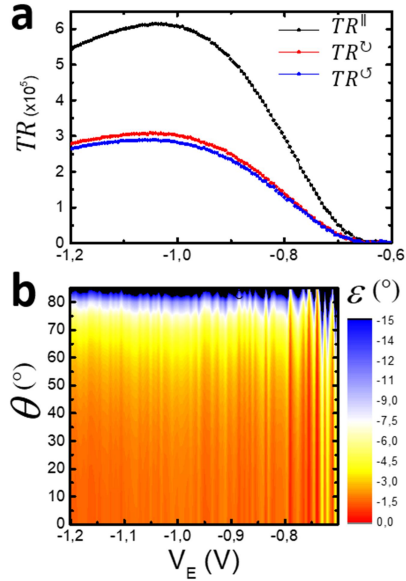

**Figure 3:** CoFeB (1 nm) precession layer. **a)** $TR^{\parallel}$ (black), $TR^{\circlearrowright}$ (red) and $TR^{\circlearrowleft}$ (blue) versus applied voltage on the tunnel barrier measured at 60 K. **b)** Precession angle calculated from TRs as a function of applied voltage on the tunnel barrier considering different filtering angles $\theta$ (measure at 60K).

Quantitative values are obtained by varying the active precession layer thickness and taking into account that electrons overcoming the Schottky barrier have $\vec{k}_{//} = \vec{0}$ (see additional material). In this case, the distance traveled by the electrons equals the thickness d of the active magnetic layer. Since the hot electron current is exponentially decreasing as a function of d (see additional material), the $sin(\varepsilon)\, cos(\theta)$ product can be rewritten as:

$$sin(\varepsilon)\, cos(\theta) \;=\; \sin(\varepsilon^* d)\, cosh(\frac{d}{2\lambda^-})^{-1}\;,$$

where $\varepsilon^*$ is the precession angle per nanometer and $\frac{1}{\lambda^-} = \frac{1}{\lambda^\downarrow} - \frac{1}{\lambda^\uparrow}$ ($\lambda^{\downarrow\uparrow}$ being the minority/majority inelastic electron mean free paths).

---

[13] In the ∥ configuration, magnetizations of the polarizer, the active layer and analyzer are parallel. Consequently $\boldsymbol{P} \cdot \boldsymbol{M_{An}} = P_0$.
[14] Magnetization Precession by Hot Spin Injection, W. Weber, S. Riesen, H. C. Siegmann, Science 291, 1015 (2001).

Fits of the $sin(\varepsilon)\, cos(\theta)$ product vs. d (reported in additional material) allows then the extraction of $\varepsilon^*$ and $\lambda^-$ as a function of hot electron energy. Precession was studied in two ferromagnetic layers, Co and CoFeB. These two materials have been chosen for the good quality of the tunnel barriers that can be achieved when the MgO is deposited on top. The thicknesses of the active layer were varied between 1 and 10 nm. As both $\varepsilon^*$ and $\lambda^-$ are mostly constant over the energy window studied, their mean values are presented in table 1 for comparison to reported values.

| | **Results comparison table** | | | |
|---|---|---|---|---|
| | **Present work** | | **W. Weber et al.** | |
| **Active Layer** | CoFeB | Co | Co | Fe |
| **Energy (above $E_F$)** | 1 eV | 1 eV | 7 eV | 7 eV |
| $\varepsilon^*$ | 2.4° nm$^{-1}$ | 0.7° nm$^{-1}$ | 19° nm$^{-1}$ | 33° nm$^{-1}$ |
| $\lambda^-$ | 0.56 nm | 1.38 nm | 1.54 nm | 1.49 nm |

**_Table :_** *Summary of the experimental values extracted from our work for Co and CoFeB and comparison to previous reports at higher energy.*

First, let's look at $\lambda^-$ for which values concerning Co can be found in the literature. All of them are in good agreement (our work, Weber et al, Van Dijken et al[15]), suggesting only a slight energy dependence of $\lambda^-$ (see additional material). These values of $\lambda^-$ indicate that Co thicknesses larger than 10 nm are necessary to almost completely turn the spin polarization vector of the hot electrons into the direction of the magnetization of the active layer. When Fe is inserted in the Co layer $\lambda^-$ decreases to 0.81 nm for $Co_{84}Fe_{16}$ in Van Dijken et al and to 0.56 nm in our work for $Co_{50}Fe_{50}$, which is linked to the decrease of $\lambda^\downarrow$. For CoFeB almost full spin filtering occurs for thicknesses of 5 nm. The fact that $\lambda^-$ does only slightly vary in our energy window has been pointed out theoretically by Nechaev et al[16]: $\lambda^-$ is mainly determined by the small value of $\lambda^\downarrow$ that does not change with energy.

The surprise of our study relies on the values of the precession angle. Since no other data are available in this low energy range ($E_F$+1eV), we can only compare with the work of Weber et al. which has been performed with the same material (Co) but at a much higher electron energy ($E_F$+7eV): the precession angle per nanometer is 30 times smaller at $E_F$+1eV than at $E_F$+7eV. By changing the active layer material to CoFeB, i.e. Co is partly replaced by Fe, we find in our lab-on-chip experiments an increase of the precession angle by a factor of 3.5. This tendency seems to be followed in the free electron beam experiments at higher electron energy when going from Co to Fe.

---

[15]S. van Dijken et al., Phys. Rev. B 66, 094417 (2002)
[16]I.A. Nechaev et al., Eur. Phys. J. B 77, 31–40 (2010)

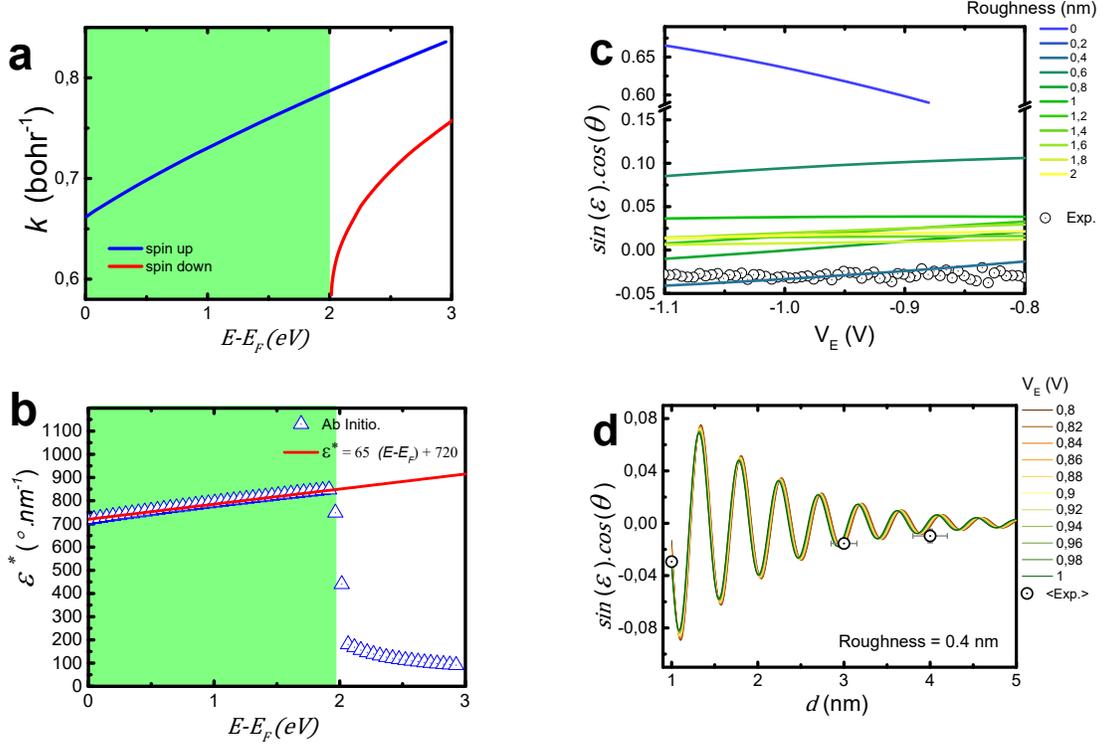

**Figure 4:** a) Ab initio computation of the band structure along Δ direction in the CoFe(100) case. b) ε* vs. E-E_F from ab initio computation in the CoFe case. c) Mean sin(ε)cos (θ) vs. V_E. Line for an active layer composed of a 1 nm thick CoFeB active layer and for various variation of travel distance. Points are experimental data. d) Mean sin(ε)cos (θ) vs. d for different values of V_E. Points : experimental values of sin(ε)cos (θ) for d= 1 nm, 3 nm, and 4 nm.

When a spin-polarized electron beam is injected into a region of space where a perpendicular magnetization exists, the precession of the spin-polarization vector is theoretically given by $\varepsilon^*(E) = \Delta k$ where $\Delta k = k^\uparrow - k^\downarrow$ is the difference in $k$-vector for both spin bands at energy E (see additional material). When only the spin-up band is accessible as for instance in figure 4a at energies below 0.7 eV we have $\Delta k = k^\uparrow$ (see additional material). As a result, the value of $\varepsilon^*$ should be strongly dependent on the spin-dependent band structure of the active layer. *Ab initio* calculations have been performed to get the band structure of CoFe(100) in the Δ direction (figure 4a) that allows the determination of $\varepsilon^*$ (figure 4b). At energies above 1.95 eV, the two spin bands are accessible and increasing the energy leads to a decrease of $\varepsilon^*$ as observed experimentally by Weber et al. Furthermore, the values are in rough agreement with the experimental report at higher energies. However, for energies below 1.95 eV, only one band can be accessed. In this case, $\varepsilon^* = k^\uparrow$ such that huge values of $\varepsilon^*$ should be measured experimentally. In real samples, however, fluctuations in the hot electron travel distance should be considered. Using the values and the linear variation of $\varepsilon^*$ from the ab initio calculations and the experimental values of $\lambda^-$, we calculated a mean value of $sin(\varepsilon)cos(\theta)$ considering a travel distance varying from $d$ to $d + \Delta d$ for a sample with $d = 1$ nm and by varying $\Delta d$:

$$sin(\varepsilon)cos(\theta) = \frac{1}{\Delta d} \int_d^{d+\Delta d} \frac{sin(\varepsilon^* \cdot t)}{cosh(\frac{t}{2\lambda^-})} \cdot dt$$

The experimental values of $sin(\varepsilon)cos(\theta)$ could be reproduced with $\Delta d$ = 0.4 nm (figure 4c). This difference in travel distance cannot be related to specular electron travelling: it would correspond to an angle of 44.4°, which is completely out of the acceptance cone. However, a layer roughness of 0.4 nm is reasonable and can be considered as being constant as a function of layer thickness. The mean values of $sin(\varepsilon)cos(\theta)$ can then be calculated as a function of $d$ and injection energy with a fixed value of $\Delta d$ = 0.4 nm. Oscillations could be calculated that are in agreement with our experimental results as seen in figure 4d. Furthermore, even if a strong variation of $\varepsilon^*$ is expected theoretically with energy, the theoretical mean values $sin(\varepsilon)cos(\theta)$ are almost constant as observed experimentally.

In conclusion, we show for the first time the manipulation of the spin direction in an unexplored energies range thanks to an all solid-state device. As forecasted theoretically, $\varepsilon^*$ is huge and requires, to be usable, a better control of the active layer roughness. This result is the starting point for new studies in which materials, crystallographic orientations, band structure are parameters that can affect precession and pave the way for exploration of far richer spin transport properties than the one at high energies.

**Contributions**

M.H. and D.L. conceived the project. C.V., M.H. and D.L. were in charge of the thin-film growth and optimization of magnetic properties. C.V. patterned the samples. Y. L. designed the lithography masks and optimized the Cu/Si Schottky barrier. C.V. and D.L. conducted the electronic transport under applied field experiment. C. T., M. C., C.V., W.W., D.L. and M.H. analyzed the data and D.L. and M.H. wrote the manuscript. All authors contributed to the discussion.

**Acknowledgments**

This work has partially been supported by ANR SpinPress (ANR-09-BLAN-0076). Experiments were carried out on IJL Project TUBE-Davm equipment funded by FEDER (EU), French PIA project "Lorraine Université d'Excellence (ANR-15-IDEX-04-LUE), Region Grand Est, Metropole Grand Nancy and ICEEL. The authors thank Sylvie Robert and Guillaume Sala for help with the X-ray diffraction and growth experiments, Stéphane Suire and Sébastien Petit-Watelot for help to set up the high field rotating field electronic transport set up and Gwladys Lengaigne for help with lithography. The authors also thank Béatrice Négulescu and Julien Bernos for their pioneering work in the laboratory on the MgO tunnel barrier optimization and Danny Petty Gweha Nyoma for help with the experiment. The authors also thank Claire Baraduc and Peter Oppeneer for fruitful discussions.

# Supplementary informations

## Emission and acceptance cones, energy filtering

1-Acceptance cone and energy filtering: the Schottky diode allows a dual analysis.

Energy filtering
On one hand, the electrons collected in the semiconductor, in our case silicon, will always have an energy larger than 0.7 eV (the height of the Schottky barrier). Thus, we will analyze the precession only of hot electrons that have an energy above 0.7 eV. All thermalized electrons will be reinjected by the tunnel barrier through the base's electron recovery circuit.

Acceptance cone
On the other hand, the conservation of the electron's wavevector parallel to the Cu/Si interface will induce the existence of an acceptance cone at the Cu/Si interface. The corresponding cone angle $\theta_c$ can be evaluated by the following equation [Vlu01]:

$$\sin(\theta_c)^2 = \frac{m^*_{Si}}{m^*_{métal}} \left( \frac{E - q\phi_B}{E + E_c} \right)$$

The value of the angle $\theta_c$ can be estimated in the case of the Cu/Si(100) interface. With $\frac{m^*_{Si}}{m^*_{Cu}}$ = 0.2, $q\phi_B$ = 0.75eV and $E_C$ = 7.1eV, we find $\theta_c$ = 4.5° at E=1eV. Thus, only electrons arriving at the Cu/Si interface at an angle to the normal at the interface of less than 4.5° can pass into the Silicon.

Experimental measurements
All the measurements reported in the paper have been performed at 60 K. Indeed, it is known that at low temperatures, less than 50K, electrons trapping effects at the Cu/Si interface have been identified in the past by our group in samples made using the same chemical process to clean the Si surface and with the same deposition conditions of the Cu layer [Lu13, Lu14]. On the other hand, at high temperatures, Schottky barrier could be leaky. In order to check those spurious effects, we have measured the transport properties of a structure that comes as close as possible to the one studied in this paper. The magneto transport-and the characteristics of the Schottky barrier have been measured at the same time in a Pt(5)Cu(20)/MgO(2.5)/Co(3)/Cu(3.5)/[Ni(0.6)/Co(0.2)]x5/Ni(0.6)/Cu(5)/Ta(1)/Cu(5)//Si[100.HF] structure as a function of temperature.

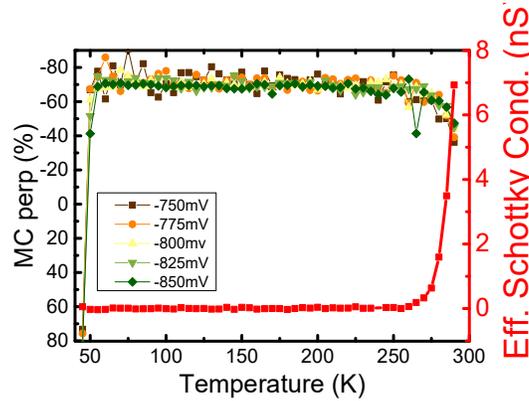

*Figure:* Variation of the crossed magneto-current of hot electrons at different energies and of the Schottky barrier conductance as a function of temperature.

$MC^\perp$ is defined as in the paper. It is the magneto-current ratio defined as $\frac{I_C^\parallel - I_C^\perp}{I_C^\perp}$ where $I_C^\parallel$ and $I_C^\perp$ are the collected currents when the magnetization of Co and Co/Ni are parallel or perpendicular to each other, respectively. The effective Schottky conductance is defined as $\frac{dI_C}{dV_E}$ measured for $V_E$ less than 100 mV. We can clearly see that for temperatures below 250 K, the effective Schottky conductance is zero and so the Schottky diode is not leaking. Above 250 K, regardless of the magnetic configuration of the spin valve, the energy barrier continues to play its role as a filter of hot electron energy but a current of thermalized electrons is added to this electron current such that the MC decreases. Thus, the measurement of the hot electrons current is no longer reliable. For temperatures less than 50K, the MC also decreases which is a sign of charge trapping as reported previously [Lu13, Lu14].

Conclusions
Thus, for an energy injection around 0.7 eV, the electrons that will be measured in the silicon will have passed through the spin valve in an almost ballistic manner. The hot electrons current measurements are reliable for temperatures between 50 and 250 K.

2-Emission cone: the tunnel junction emits electrons with k-vector mainly perpendicular to the interface

As it will be discussed in the next section, when electrons are injected from a magnetic tunnel junction [Man08], the probability of injection depends on the k-vector direction. In the simplest model, the highest injection probability occurs when the k-vector is perpendicular to the barrier interface. As a result, k-vector selection is done by the tunnel barrier [Man08] and an emission cone exists.

References :

[Vlu01] Modeling of spin-dependent hot-electron transport in the spin-valve transistor, R. Vlutters, O. M. J. van 't Erve, R. Jansen, S. D. Kim, J. C. Lodder, A. Vedyayev, B. Dieny, Phys Rev B 65, 024416 (2001)
[Lu13] Interfacial Trapping for Hot Electron Injection in Silicon,Y. Lu, D. Lacour, G. Lengaigne, S. Le Gall, S. Suire, F. Montaigne and M. Hehn, Applied Phys. Lett. 103, 022407 (2013).
[Lu14] Electrical control of interfacial trapping for magnetic tunnel transistor on silicon, Y. Lu, D. Lacour, G. Lengaigne, S. Le Gall, S. Suire, F. Montaigne, M. Hehn, M. W. Wu, Appl. Phys. Lett. 104, 042408 (2014).
[Man08] A. Manchon, N. Ryzhanova, A. Vedyayev, M. Chschiev and B. Dieny, Description of current-driven torques in magnetic tunnel junctions, J. Phys.: Condens. Matter 20, 145208 (2008).

## Discussion on the $k_{//} = 0$ hypothesis

It is well known in solid state physics that when an electron moves through a material, it interacts not only with other electrons in the material but also with pseudoparticles such as phonons and magnons. The higher the energy of the electrons, the more electron-electron interaction dominates scattering processes. This makes an essential difference between our lab-on-chip experiments and the free-electron beam experiments of W. Weber et al: in our measurements, for electrons with low energies of around $E_F$+1eV, all types of interaction have to be considered.

This is shown in following figure. Nechaev's calculations show that the mean free path of spin up electrons is overestimated with respect to experimental values (blue triangles) at low energies if only electron-electron interaction is taken into account ([Nec10], dark blue curve in the figure). By including an interaction with pseudoparticles (phonons and/or magnons) with a typical energy of 50 meV, the mean free path values are close to those measured experimentally (cyan blue curve in the figure). In the case of spin down electrons (red line and triangles) the agreement between theory and experiment is already quite good without the inclusion of the interaction of the electrons with pseudoparticles. In fact, for spin down electrons the electron-electron interaction is so strong that it overwhelms completely other possible types of interaction.

Thus, electron-electron, electron-phonon and electron-magnon interactions have to be considered in general. The question is now to what extent these interactions can modify the electron trajectory and in particular the distance travelled by the electrons in the active layer. In any case, the relevant interaction, i.e. the interaction that influences the electrons which pass the Schottky barrier, must be elastic or quasi-elastic (energy filtering of the Schottky barrier) and the electrons must impact the Schottky barrier with a small $\vec{k}_{//}$ to be within the acceptance cone with a cone angle of 4,2°.

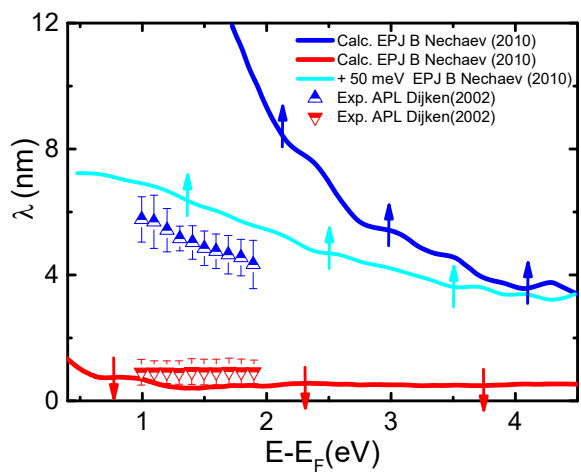

***Figure:*** *Variation of the mean free path of hot electrons with energy calculated by Nechaev et al. (Eur. Phys. J. B 77, 31–40 (2010)). Blue curve: mean free path of spin up electrons by only considering electron-electron interaction; Red curve : mean free path of spin down electrons by only considering electron-electron interaction; Blue cyan curve: mean free path of spin up electrons by adding an interaction with a particle with 50meV energy; Blue triangles: mean free path of spin up electrons measured experimentally; Red triangles: mean free path of spin down electrons measured experimentally.*

- electron-electron interaction

For electron-electron interaction processes, Fermi's liquid theory predicts that a hot electron will lose half of its energy, the other half allowing a Fermi-level electron to become a hot one. Thus, this interaction is extremely inelastic and electrons that have undergone such an interaction will not be collected in the Si and will thus not participate in the detected current. As a result, electron-electron interaction exists in the range of energy but the electrons that experience an interaction with another electron will not contribute to the hot electron current.

- electron-phonon interaction

The process of electron-phonon interaction is indeed quasi-elastic but leads to electronic backscattering because of the strong wavevector change. Thus, the electron returns to its emission zone, i.e. the tunnel junction. It will be reflected there and will return towards the Schottky barrier. Thus, electrons subjected to an interaction with phonons will pass at least 3 times through the active precession layer. This will greatly increase their chance of losing energy (due to electron-electron interaction) and not being collected in the Si. We will thus consider that electron-phonon interaction is not relevant for the detected current.

- electron-magnon interaction

The electron-magnon interaction process is also quasi-elastic but leads to a spin flip of the electron spin. When a spin-flip occurs in a layer where the magnetization is perpendicular to the spin, the spin-flip process leads to a reorientation of the spin along the magnetization direction and thus to an additional spin-filtering process. However, when we compared our values for the spin filtering angle with those measured at higher energies, we did not notice a big difference. Thus, electron-magnon scattering is not believed to be relevant in our experiments.

In conclusion, none of the hot electron interaction processes alone seems to explain the small precession angle values estimated using our measurements. We can then evoke classical decoherence effects of our electron beam.

Classical decoherence

The depth dependence of absorption of a spin current has been predicted for transverse spin currents in ferromagnetic layers. It arises from the transverse spin coherence length, which depends upon the Fermi surface integration. In spinvalve structures, for which integration has to be done over all k-vectors, the length scale is set to first order by [Zha04, Zwi05, Sti02, Car07, Wan08, Pet12]

$$\lambda_J \sim \pi/|k^\uparrow - k^\downarrow|$$

Theoretical estimations of this length scale in spinvalve structures yield 1 to 2nm, values that have also been confirmed experimentally [Gho12]. As a result, the precession angle should average to zero if the thickness of the magnetic layer is larger than 2 nm (case a in next figure).

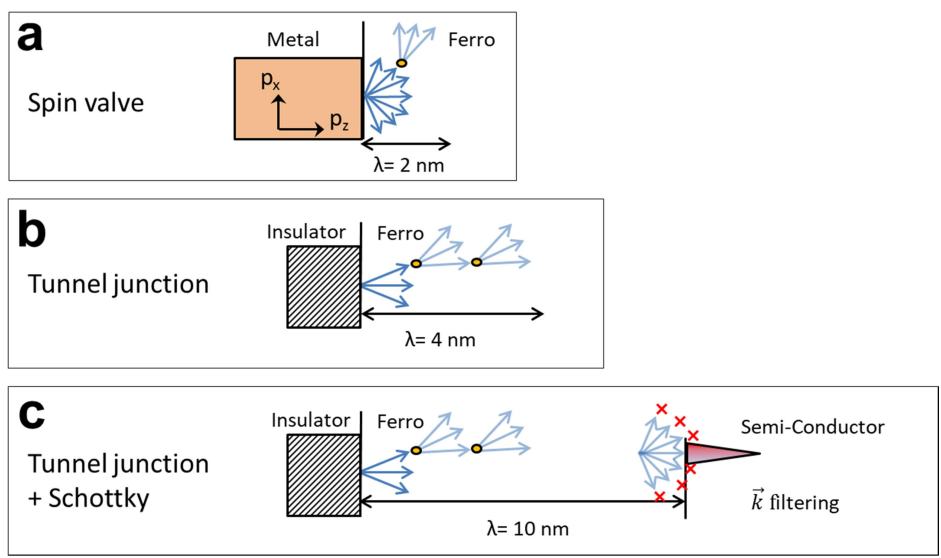

*Figure:* Classical decoherence evoked for different spintonics devices in the literature.

Things appear to be different when the electrons are injected from a magnetic tunnel junction ([Man08], [Chs15], case b in the figure). In this case, the transverse component of the spin density is damped by 50% within the first few nanometers. This decay length, evaluated to be 4 nm, is very large compared to previous theoretical predictions and experimental investigations on spinvalves. In the case of tunnel junctions, the averaging of torques will be less destructive than in metallic spin valves where all the Fermi surface is involved in the quantum interferences. This arises from the fact that $\lambda_J$ has not to be integrated over all the k-vector direction since k-vector selection is done by the tunnel barrier [Man08].

In our case (case c in the figure), the injection of hot electrons is the one described in figure b but, though the use of the collecting Schottky barrier, the integration has to be done for k-vectors in the acceptance cone $\theta_C = 4.5°$. The decay length over which average precession angle is zero is much longer than 4nm. Our experimental results suggest that it should be longer than 10nm.

Experimental measurements

From the experimental point of view, it is very difficult to test the existence of hot electron diffusion. For a given device, we can play with the temperature to activate diffusion and see if the collected current is strongly dependent on temperature. We have to keep in mind that temperatures only between 60 and 250 K are allowed to avoid spurious effects from the Schottky barrier (see section Emission and acceptance cones, energy filtering). Therefore, the transfer ratio has been measured as a function of temperature and injection energy in the configuration for which all the magnetizations are parallel. As can be seen in the following figure, the TR shows a variation of only 4% when the temperature varies from 60 to 100 K. Diffusion of the collected electrons is thus very limited.

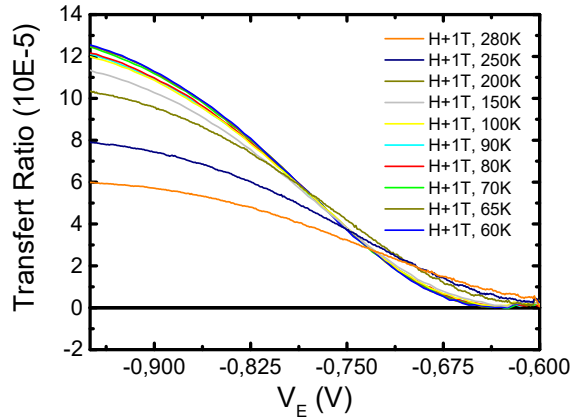

*Figure :* Variation of the transfer ratio in the parallel configuration as a function of injection energy and temperature

References :
[Nec10] I.A. Nechaev et al., Eur. Phys. J. B 77, 31–40 (2010)
[Zha04] J. Zhang, P. Levy, S. Zhang, and V. Antropov, Phys. Rev. Lett. 93, 256602 (2004).
[Zwi05] M. Zwierzycki, Y. Tserkovnyak, P.J. Kelly, A. Brataas, and G.E.W. Bauer, Phys. Rev. B 71, 064420 (2005).
[Sti02] M.D. Stiles and A. Zangwill, Phys. Rev. B 66, 014407 (2002).
[Car07] K. Carva and I. Turek, Phys. Rev. B 76, 104409 (2007).
[Wan08] S. Wang, Y. Xu, and K. Xia, Phys. Rev. B 77, 184430 (2008).
[Pet12] C. Petitjean, D. Luc and X. Waintal, Phys. Rev. Lett. 109, 117204 (2012).
[Gho12] A.Ghosh, S. Auffret, U. Ebels, and W.E. Bailey, Penetration Depth of Transverse Spin Current in Ultrathin Ferromagnets, Phys. Rev. Lett. 109, 127202 (2012).
[Man08] A. Manchon, N. Ryzhanova, A. Vedyayev, M. Chschiev and B. Dieny, Description of current-driven torques in magnetic tunnel junctions, J. Phys.: Condens. Matter 20, 145208 (2008)
[Chs15] M. Chshiev, A. Manchon, A. Kalitsov, N. Ryzhanova, A. Vedyayev, N. Strelkov, W. H. Butler, and B. Dieny, Phys. Rev. B 92, 104422 (2015).

## Variation of filtering with thickness

As expected and reported in previous studies, the hot electron current decreases exponentially with the active layer thickness

$$I^{\uparrow(\downarrow)} \propto e^{-d/\lambda^{\uparrow(\downarrow)}}$$

where $\lambda^{\uparrow(\downarrow)}$ are the thermalization length for spin up and spin down hot electrons. We have measured the transfer ratios (TR) as a function of precession layer thickness (in the following figure X=CoFeB) in the parallel configuration. As shown in figure X, ln(TR) varies linearly with CoFeB thickness.

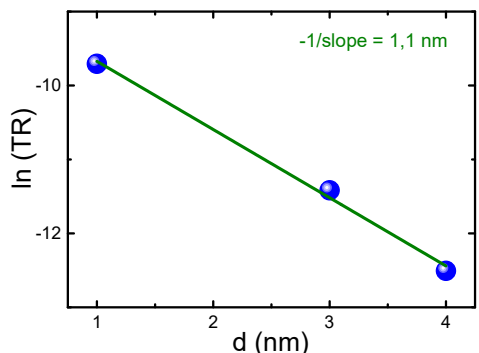

**_Figure:_** *Variation of the logarithm of the transmission ratio in the parallel configuration as a function of the active layer thickness in the case of CoFeB.*

From this experimental result, we can rewrite the expression of the spin asymmetry as:

$$A = \frac{I^\uparrow - I^\downarrow}{I^\uparrow + I^\downarrow} = \tanh\left(\frac{d}{2\lambda^-}\right) \text{ with } \frac{1}{\lambda^-} = \frac{1}{\lambda^\downarrow} - \frac{1}{\lambda^\uparrow}.$$

The spin polarization after precession can then be expressed as:

$$\vec{P} = \begin{pmatrix} P_0 \cos(\theta)\cos(\varepsilon) \\ P_0 \cos(\theta)\sin(\varepsilon) \\ P_0\sqrt{1-\cos^2(\theta)} \end{pmatrix} = \begin{pmatrix} P_0\sqrt{1-A^2}\cos(\varepsilon) \\ P_0\sqrt{1-A^2}\sin(\varepsilon) \\ P_0 A \end{pmatrix} = \begin{pmatrix} P_0\cos(\varepsilon)\cosh\left(\frac{d}{2\lambda^-}\right)^{-1} \\ P_0\sin(\varepsilon)\cosh\left(\frac{d}{2\lambda^-}\right)^{-1} \\ P_0\tanh\left(\frac{d}{2\lambda^-}\right) \end{pmatrix}.$$

As a result, the experimental results are fitted by the relation:
$$\frac{TR^\circlearrowright - TR^\circlearrowleft}{2T^{\parallel} - (TR^\circlearrowright + TR^\circlearrowleft)} = \sin(\varepsilon)\cosh\left(\frac{d}{2\lambda^-}\right)^{-1}.$$

## Variation of filtering with thickness

<u>Cobalt-Iron alloy</u>
Since the values of $\lambda^{\uparrow(\downarrow)}$ for CoFeB could not be found in the literature, in a first step we made the assumption that $\varepsilon = \varepsilon^* d$. As shown in the following figure a and b, values of $\varepsilon^*$ and $\lambda^-$ could be extracted. The results of the fits show that the quantities are constant over the range of energy studied. In a second step, no assumption has been made on the thickness variation of $\varepsilon$. We find again a linear variation of $\varepsilon$ with active layer thickness.

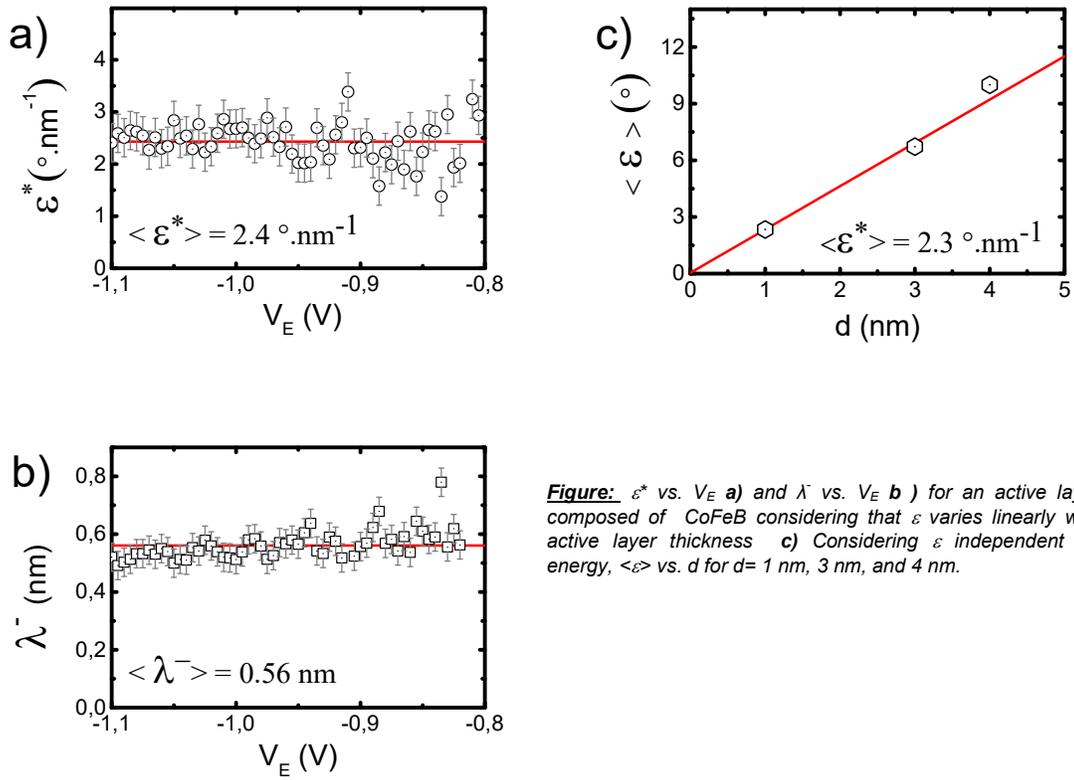

*Figure:* $\varepsilon^*$ vs. $V_E$ **a)** and $\lambda^-$ vs. $V_E$ **b )** for an active layer composed of CoFeB considering that $\varepsilon$ varies linearly with active layer thickness **c)** Considering $\varepsilon$ independent on energy, $<\varepsilon>$ vs. d for d= 1 nm, 3 nm, and 4 nm.

Cobalt

No assumption has been made on the thickness variation of $\varepsilon$ with d. We find a linear variation of $\varepsilon$ with active layer thickness. The values of $\lambda^{\uparrow(\downarrow)}$ are in agreement with literature. The results of the fits show that the quantities are constant over the range of energy studied.

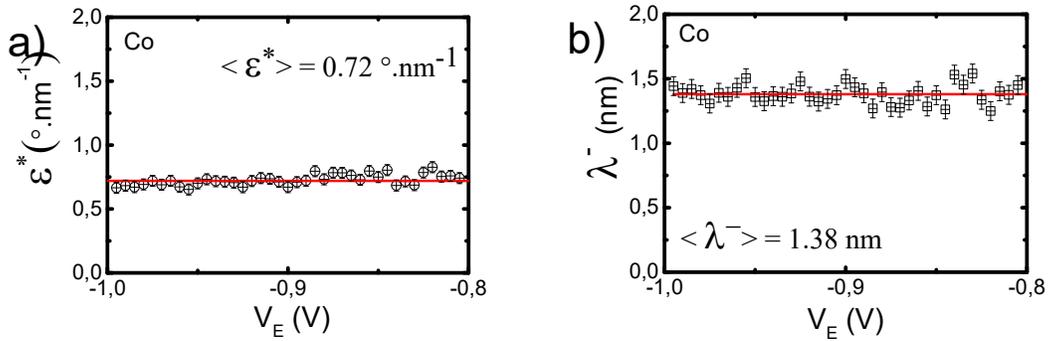

*Figure:* $\varepsilon^*$ vs. $V_E$ **a)** and $\lambda^-$ vs. $V_E$ **b )** for an active layer composed of Co

.

## Precession in a one band structure

The quantum state of an electron in the precession layer is given by:

- The following expression when two bands of opposite spin can be accessed with wave vectors $k_+$ et $k_-$ ($\theta$ is the filtering angle).

$$|\Psi> = \cos\left(\frac{\theta}{2}\right)e^{ik_+x}|+> +\sin\left(\frac{\theta}{2}\right)e^{ik_-x}|-> .$$

The average of the spin components along x and y are then expressed as:

$$<S_x> = \frac{h}{4\pi}\sin\theta \cos(\Delta k. x) ,$$
$$<S_y> = \frac{h}{4\pi}\sin\theta \sin(\Delta k. x) .$$

The precession angle per nanometer is given by $\Delta k = k_+ - k_-$ .

- The following expression when only one band of spin (here spin up) can be accessed with wave vector $k_+$. Electron waves of the other spin direction (here spin down), however, can only exist as evanescent waves with a typical decay length $1/K$.

$$|\Psi> = \cos\left(\frac{\theta}{2}\right)e^{ik_+x}|+> +\sin\left(\frac{\theta}{2}\right)e^{-Kx}|->$$

The average of the spin components along x and y are thus expressed as:

$$<S_x> = \frac{h}{4\pi}\sin\theta \cos(k_+. x)e^{-K} ,$$
$$<S_y> = \frac{h}{4\pi}\sin\theta \sin(k_+. x)e^{-Kx} .$$

Consequently, the precession angle per nanometer is given by $k_+$ .

## Co50Fe50 band structure

*Ab initio* calculations have been performed to get the band structure of CoFe(100) in the $\Delta$ direction and symmetry analysis led to identify the $\Delta_1$, $\Delta_2$ and $\Delta_5$ bands for spin up and spin down electrons. At energies above 1.95 eV, $\Delta_1$ spin up and down bands are accessible. However, for energies below 1.95 eV, only $\Delta_1$ spin up band can be accessed.

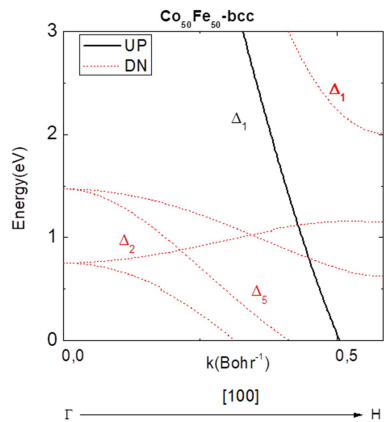

***Figure:*** *Band structure of CoFe(100) in the $\Delta$ direction.*

## Methods

Samples were grown by Ultra High Vacuum sputtering on a fluorhydric acid desoxidized low resistance Si substrate. The base pressure was lower than $5\times10^{-9}$ mbar and pure Ar at pressure of $5\times10^{-3}$ mbar was used to sputter the targets.

After multilayer deposition and thermal annealing under applied field to set the exchange bias, 4 steps of lithography are performed to make electrical contacts on the polarizer layer, on the active layer/analyzer layers and on the Si. In the following figure, the details of the contacts are given and their spatial location on the device.

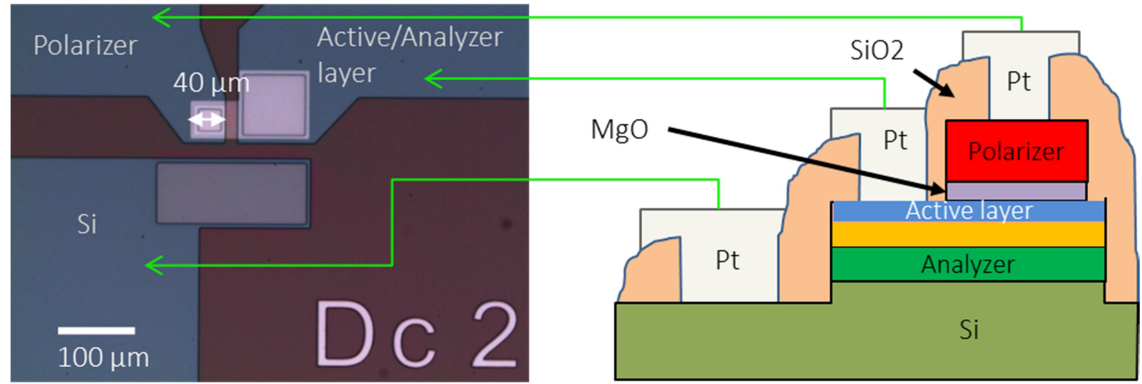

*Figure :* Device used to measure the spin precession in an all solid state device to access to low energy hot electron production. On the left, an optical microscopy picture of the contacts on the surface. On the right, a drawing of the cross section of the device with the different electrical contacts.

All the details can be found in previous papers published on the hot electron transport of crossed magnetization MTT (Magnetic tunnel transistor with a perpendicular Co/Ni multilayer sputtered on Si/Cu(100) Schottky diode, C. Vautrin, Y. Lu, S. Robert, G. Sala, O. Lenoble, S. Petit-Watelot, X. Devaux, F. Montaigne, D. Lacour, M. Hehn, J. of Phys. D : Appl. Phys. 49, 355003 (2016)) or on the angular dependence of magnetocurrent of hot electrons in MTT (Thickness and angular dependence of magnetocurrent of hot electrons in magnetic tunnel transistor with crossed anisotropies,C. Vautrin, D. Lacour, G. Sala, Y. Lu, F. Montaigne, M. Hehn, Phys. Rev. B 96, 174426 (2017)).

**Rough measurements done on the MTT**

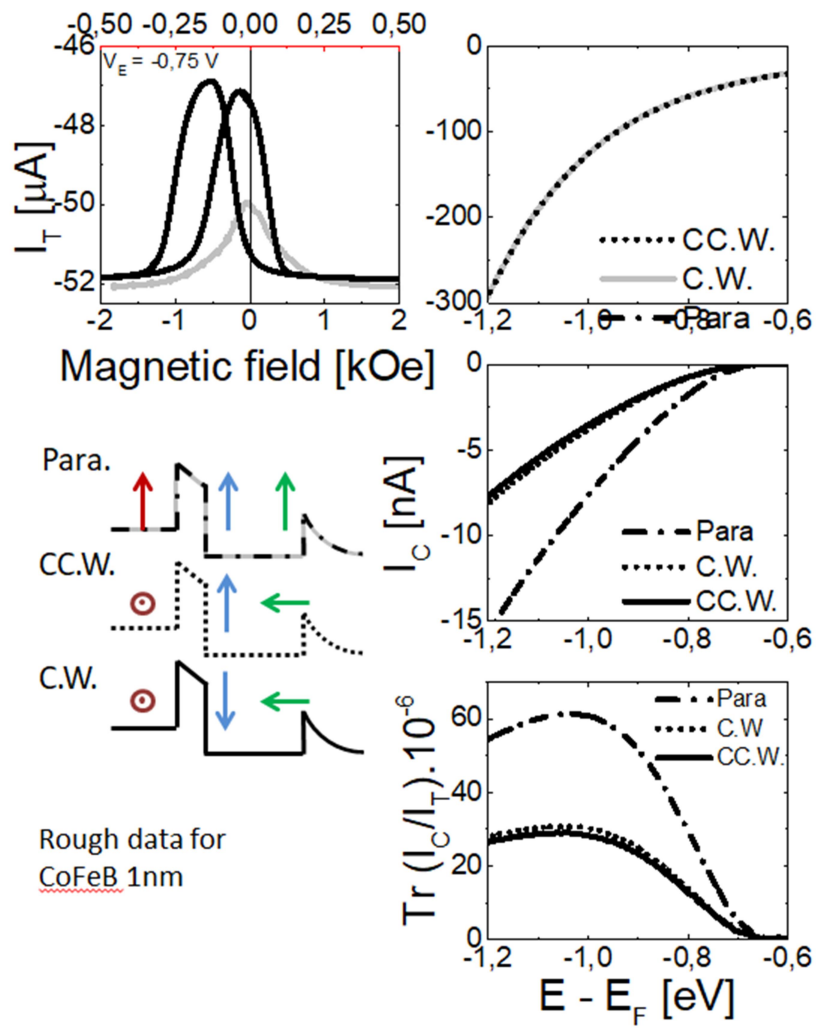

Rough data for CoFeB 1nm

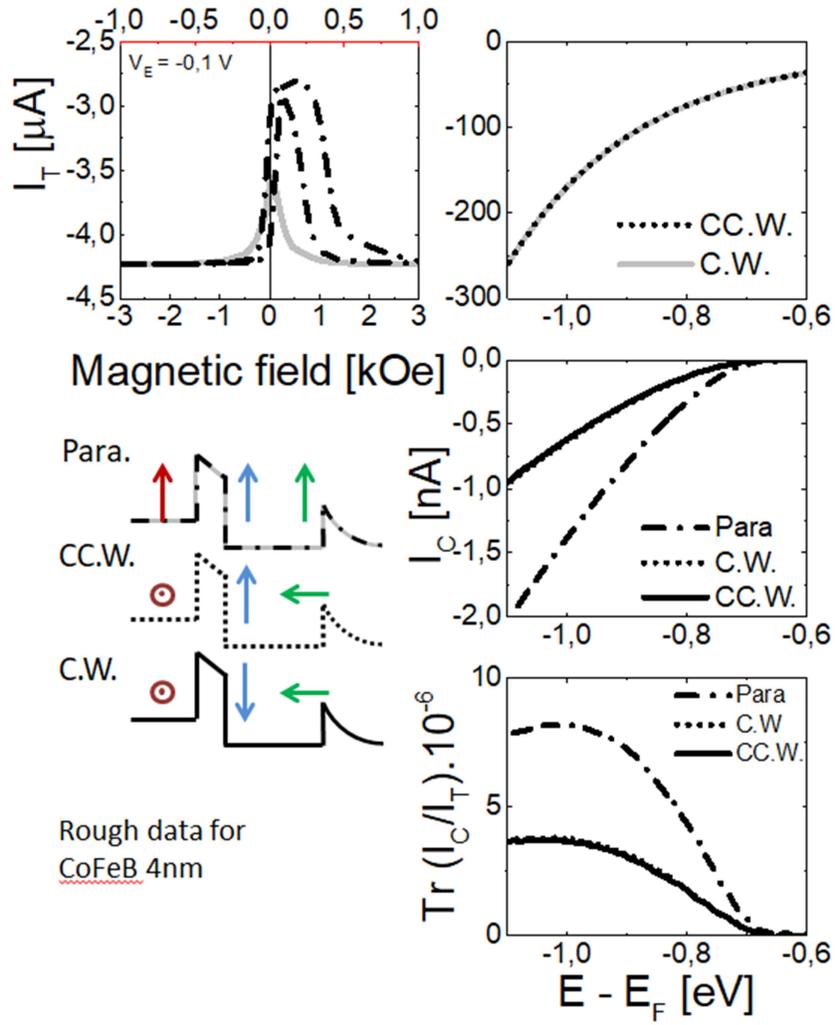

Rough data for CoFeB 4nm

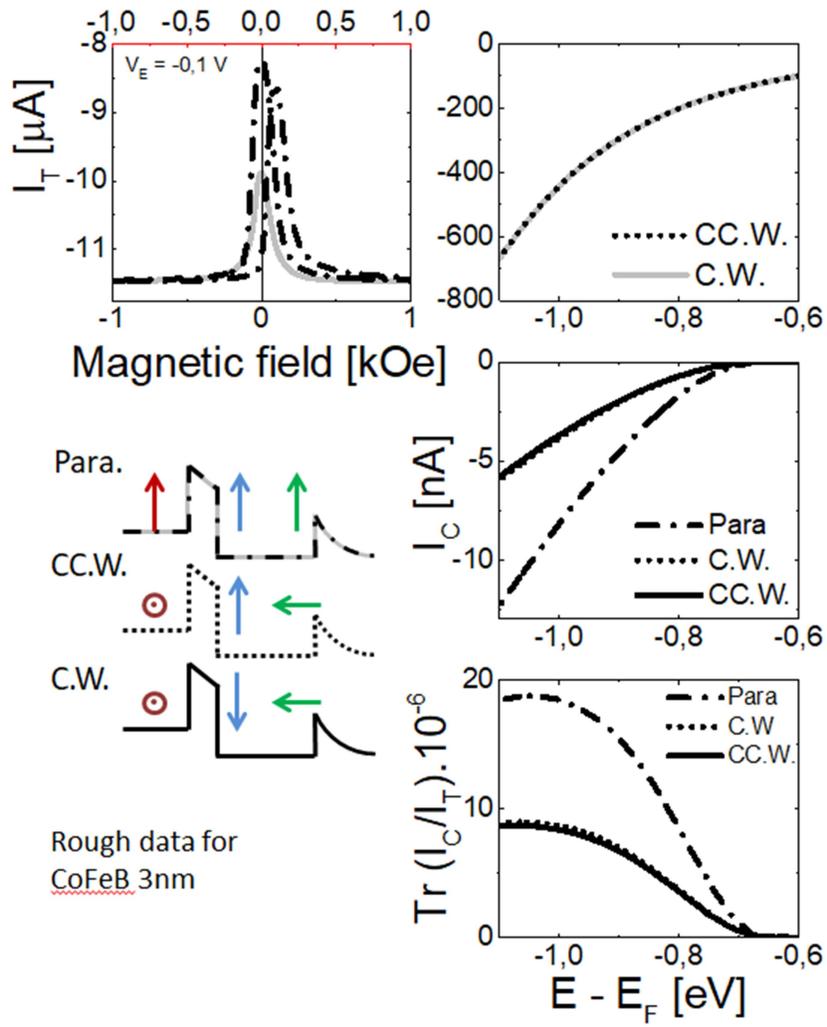

Rough data for CoFeB 3nm

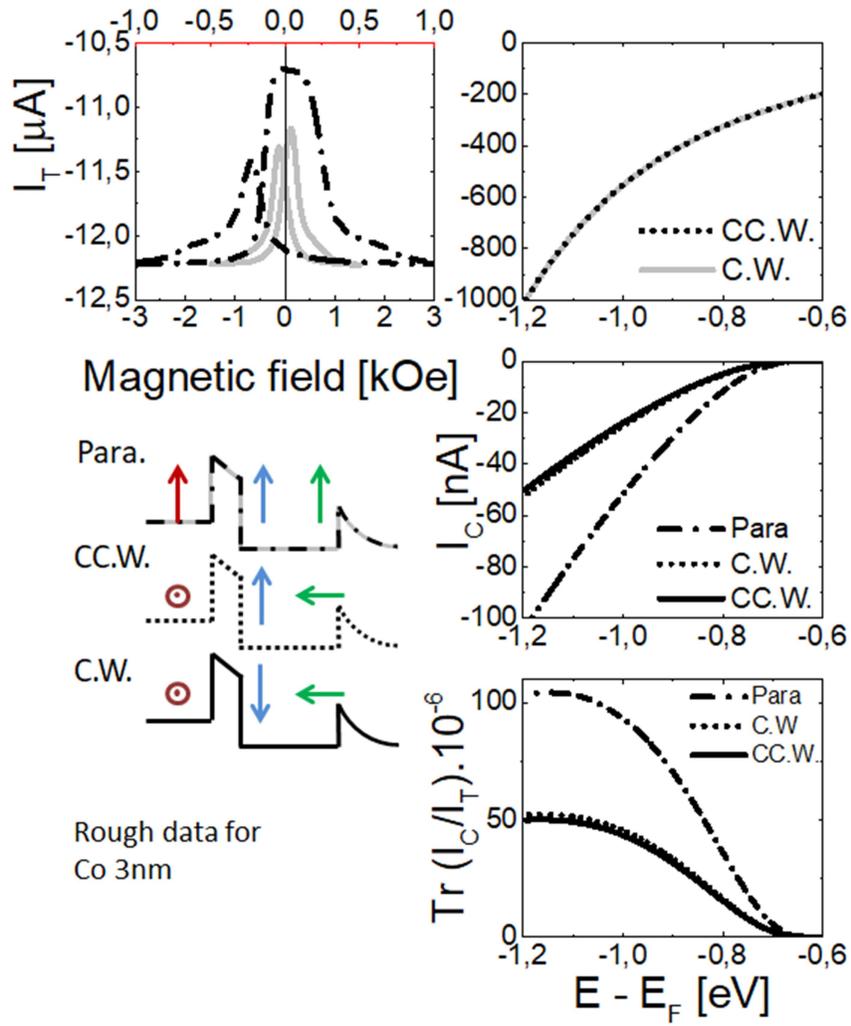

Rough data for Co 3nm

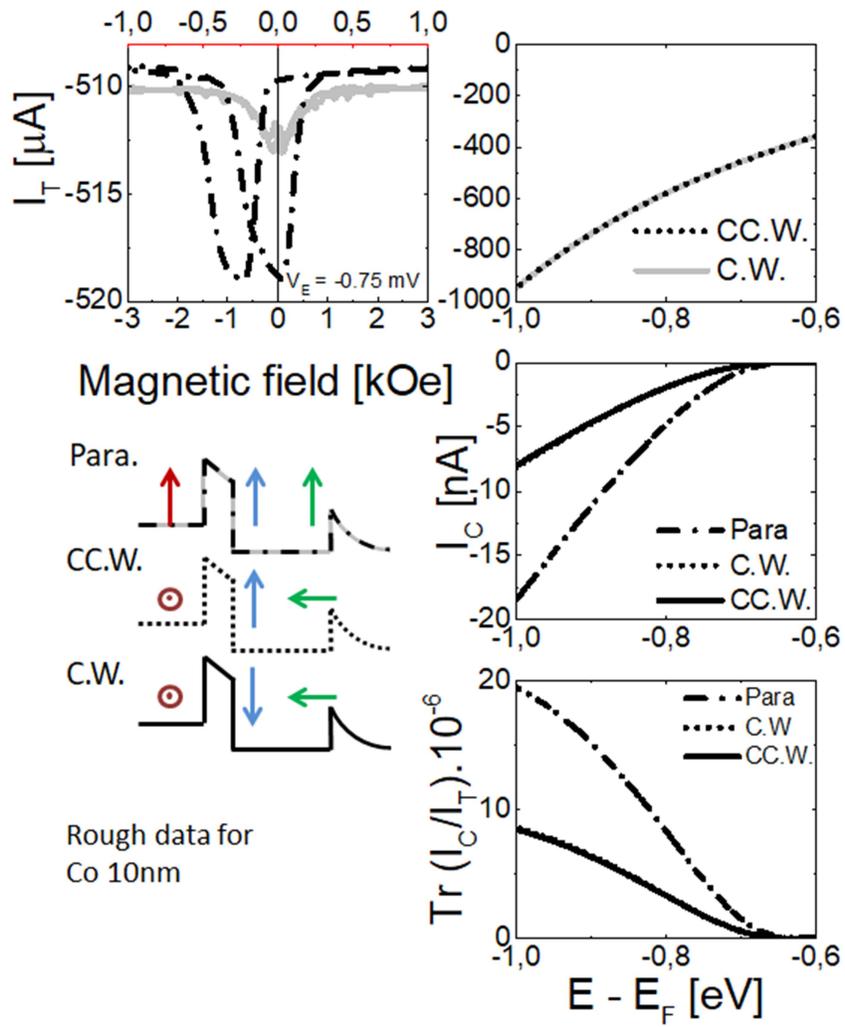

Rough data for Co 10nm